# The conducting spin-crossover compound combining Fe(II) cation complex with TCNQ in a fractional reduction state

Yuri N. Shvachko,[a,*] Denis V. Starichenko,[a,*] Aleksander V. Korolyov,[a] Eduard .B. Yagubskii,[b,*] Alexander I. Kotov,[b] Lev I. Buravov,[b] Konstantin A. Lyssenko,[c] Vladimir N Zverev,[d, e] Olga G. Shakirova,[f] Ludmila G. Lavrenova [g]

[a] *M.N. Miheev Institute of Metal Physics, Ural Branch of Russian Academy of Sciences, S. Kovalevskaya str., 18, Yekaterinburg 620137, Russia.*
[b] *Institute of Problems of Chemical Physics, Russian Academy of Sciences, Semenov ave., 1, Chernogolovka 142432, MD, Russia.*
[c] *A.N. Nesmeyanov Institute of Organoelement Compounds, Russian Academy of Sciences, Vavilov str., 28, Moscow 119991, Russia*
[d] *Institute of Solid State Physics, Russian Academy of Sciences, Ossipyan street, 2, Chernogolovka 142432, MD, Russia*
[e] *Moscow Institute of Physics and Technology, Dolgoprudnyi, Moscow region, 141700, Russia*
[f] *Komsomolsk-on-Amur State Technical University, Lenin ave., 27, Komsomolsk-on-Amur 681013, Russia*
[g] *Nikolaev Institute of Inorganic Chemistry, Siberian Branch of Russian Academy of Sciences, Lavrentyev ave., 3, Novosibirsk 630090, Russia*

**ABSTRACT:** The radical anion salt [Fe{HC(pz)$_3$}$_2$](TCNQ)$_3$ demonstrating conductivity and spin-crossover transition (SCO) associated with Fe(II) complex cation subsystem has been synthesized and structurally characterized. A pronounced change of the dc resistivity and a robust broadening of the EPR signal triggered by the SCO transition are found. Enhanced EPR line broadening is interpreted in terms of cross relaxation between the TCNQ and Fe(II) spin subsystems.

## INTRODUCTION

Molecular materials, in which conductivity and magnetism cooperate in the same crystal lattice, attract attention, because their synergy may lead to novel physical phenomena.[1] So far, the basic class of such materials was mainly represented by the quasi-two-dimensional (super)conductors based on the radical cation salts of bis(ethylenedithio)tetrathiafulvalene (BEDT-TTF) and its derivatives with paramagnetic metal complex anions of different nature.[1b-f] It is of a considerable interest to use the octahedral cation complexes of Fe(II) and Fe(III), showing reversible spin-crossover (SCO) between high-spin (HS) and low-spin (LS) states of the metal ion, as a magnetic subsystem in combination with an anion conducting subsystem.[2] The conducting sublattice of such hybrid materials could be represented by the radical anion subsystem based on [M(dmit)$_2$]$^{\delta-}$ complexes (M=Ni, Pd, Pt; dmit = 4,5-dithiolato-1,3-dithiole-2-thione; $0 < \delta < 1$)[3] and/or 7,7,8,8,-tetracyanoquinodimethane ((TCNQ)$^{\delta-}$, $0 < \delta < 1$)[4] in a fractional oxidation or reduction states, respectively. The SCO induced by temperature, pressure or light irradiation is accompanied by the changes in the coordination environment of the metal ion.[5] The electrical conductivity of most molecular conductors is very sensitive to external and/or chemical pressure.[6] There is every reason to believe that spin-crossover transition would affect the conductivity at least via a chemical compression. Furthermore, magnetic interactions between the subsystems make possible a realization of a spin-dependent electronic transport. Several conducting spin-crossover systems are known that combine the cation complexes of Fe(III) with [M(dmit)$_2$]$^{\delta-}$ anions[7] and the only system containing TCNQ$^{\delta-}$ anion.[8] The interplay between spin-crossover and conductivity was reported for two of them.[7e,f] Although the variations of the spin moment at SCO are the same for Fe(II) and Fe(III), $\Delta S = 2$, the average bond length for a given donor-atom changes significantly greater in the case for Fe(II) ions ($\Delta r \approx 8-15\%$) rather than that for Fe(III) ions ($\Delta r \approx 5\%$). Besides Fe(II) complexes are diamagnetic in LS and paramagnetic in HS states. In this context, the Fe(II) complexes seem preferable for the design of conducting SCO systems. The synthesis of electroactive SCO systems based on a redox labile Fe(II) complexes with [M(dmit)$_2$]$^{\delta-}$ anions is most likely impossible, principally because of high oxidation potential of [M(dmit)$_2$]$^{-}$ to [M(dmit)$_2$]$^0$. Recently, Nihei et al. has proposed an alternative approach to design the conducting SCO compounds based on Fe(II) complexes: the ligands contain potentially conducting fragments such as tetrathiafulvalene (TTF) and its derivatives.[9] The SCO complex with the conductivity $\sigma_{rt} = 2.6 \times 10^{-3}$ $\Omega^{-1}$ cm$^{-1}$ and coupling between transport properties and the spin transition was synthesized within this approach. The domain of the conducting TCNQ radical anion salts with SCO Fe(II) cation complex as counterion remains essentially unexplored. The only structurally characterized SCO Fe(II) complex with TCNQ$^{.-}$ counterion, [Fe(abpt)$_2$(TCNQ)$_2$], is known up to date.[10] However this complex is an insulator ($\sigma_{rt} = 5 \times 10^{-10}$ $\Omega^{-1}$ cm$^{-1}$).[10b] Dielectric properties are associated primarily with strong dimerization of TCNQ radicals and absence of long-range stacking. More recently the materials [Fe$^{II}$(tpma)(xbim)](X)(TCNQ)$_{1.5}$·DMF (X = ClO$_4$, BF$_4$), [Co$^{II}$(terpy)$_2$](TCNQ)$_3$·CH$_3$CN and [Co$^{II}$(pyterpy)$_2$](TCNQ)$_2$ showing SCO and electrical conductivity have been obtained.[11] Here, the first three compounds have a relatively high conductivity ($\sigma_{rt} \sim 2 \times 10^{-1}$ $\Omega^{-1}$cm$^{-1}$) and co-exist with a partially reduced TCNQ$^{\delta-}$ radical ($\delta = 0.67$), whereas the latter compound reveals low conductivity ($\sigma_{rt} = 5 \times 10^{-6}$ $\Omega^{-1}$cm$^{-1}$) and coexists with a fully reduced TCNQ$^{1-}$ radical.

In this paper, we report synthesis, structure, and physical properties of the new conducting spin-crossover compound based on the Fe(II) cationic complex with TCNQ counterion in a fraction reduction state, [Fe{HC(pz)$_3$}$_2$](TCNQ)$_3$ (**I**), where HC(pz)$_3$ = tris(pyrazol-1-yl)methane. The dc-resistivity and the linewidth of the anion radical EPR signal undergo robust changes associated with the SCO transition. The interactions between the conducting spin system and local moments of Fe(II) ions can possibly contribute to the observed effect.

**RESULTS AND DISCUSSION**

**Synthesis and structure**

Black rhombic crystals of **I** were obtained in argon atmosphere by spontaneous crystallization from solution containing [Fe{HC(pz)$_3$}$_2$]I$_2$·H$_2$O[12] and TCNQ (see Experimental). The synthesis was based on ability of a neutral TCNQ molecule to be reduced by an iodide ion to a radical anion.[4a] Thermal analysis of **I** has shown that the complex is stable up to 200 ºC. The DSC curve reveals the exothermal peaks at 251.3 and 272.1ºC, associated with decomposition of the complex. As this takes place, the ions with *m/e* 26 (CN) are observed in the mass spectrum (Supporting information, Fig. S1).

X-ray structural analysis at 100 K indicates that **I** crystallizes in the triclinic space group P-1 (Z = 1, Z'= 0.5) with three TCNQ molecules per one [Fe{HC(pz)$_3$}$_2$]$^{2+}$ complex cation (Fig. 1). The cation and one of the TCNQ molecules occupy special positions, i.e. center of symmetry; so there are only two independent TCNQ molecules, further denoted as **A** (one in the general position) and **B** (one at the center of symmetry). The overall crystal structure of **I** can be described as that consisting of cationic and anionic layers parallel to the crystallographic plane *ab* (Fig. 2). In the [Fe{HC(pz)$_3$}$_2$], the Fe(II) ion at 100 K must be in a low-spin state, as judging by the average Fe-N bond length (1.968(1) Å) (Supporting information, Fig S2, Table S1). Indeed, the value of ca 1.97 Å is typical for the low-spin (LS) octahedral FeN$_6$ complexes, whereas in a high-spin (HS) state of the FeN$_6$ complexes the average Fe-N bond distance is 2.17 Å.[13] All other parameters, including the average torsion angle Fe(1)-N-N-C (177.63°, Table S1) also coincide with the expected values for the LS state.[13b,c] The stacking-bonded cationic chains run along the crystallographic axis *a* while the TCNQ anions are stacked along the [110] direction (Supporting information, Fig S3). The cation…cation stacking interaction is observed between the independent pyrazole cycles (C…C contacts are 3.343(2) Å), (Supporting information, Fig. S4). In the anionic stacks, centrosymmetric triads **A…B…A** are formed with a rather strong stacking interaction (Fig. 3). The interplane **A…B** separation is 3.05 Å, and the distance for the shortest C…C contacts varies in the range of 3.101(4) – 3.328(2) Å. The above triads are linked by **A…A** stacking interactions (the interplane distance is 3.33 Å) mainly involving C(CN)$_2$ groups (Supporting information, Fig. S5). These stacks are assembled into layers by C…N interactions and weak side-by-side C-H…N-C hydrogen bonds (Supporting information, Fig. S6, S7). Cation…anion interactions in **I** also include weak C-H…N contacts (Fig. 2). The majority of these contacts belong to the molecule TCNQ (**B**). Based on the formula of **I**, the average charge on the TCNQ$^{\delta-}$ radical should be -0.67 or -2.0 on repeating **ABA** triad of the stack. It was shown that the charge state of TCNQ could be estimated from its bond distances.[14] There is a great deal of work relating the carbon-carbon bond lengths in TCNQ to the degree of charge ($\delta$) held on it.[14a,b] By using the Kistenmacher relationship,[14b-d] $\delta$ = A[c/(b+d)] + B (A = -41.667, B = 19.833, c, b, d are the average bond lengths of TCNQ shown in Scheme, Tables 1 and S2), we estimated the charge distribution for **A** and **B** TCNQ molecules in the stack (-0.5 and -0.89, respectively, Table 1). These values point to different degrees of a charge acceptance by the TCNQ species in **I**. The total charge on the various TCNQ species is -1.89, which is in agreement with the +2 charge on the [Fe{HC(pz)$_3$}$_2$] cation.

**Conductivity and magnetic properties**

The normalized dc resistance, R(T)/R(300 K), measured by a standard four-probe method for the single crystal [Fe{HC(pz)$_3$}$_2$]$^{2+}$(TCNQ)$_3^{2-}$ demonstrates a semiconducting type behavior (Fig. 4). The placement of the electrodes on the crystal is shown in Fig. S8. The value of the conductivity at room temperature is $1.5 \times 10^{-2}$ $\Omega^{-1}$cm$^{-1}$. Below 280 K the datapoints are well described by the exponential law $R_{calc}(T) = \exp(\Delta E/kT)$ with the energy gap $\Delta E$ = 0.17 eV (Inset in Fig. 4). The results of the Montgomery method measurements (see Experimental) above 200 K are presented in the Figs. 5 and 6 for the in-plane ($\rho_\parallel$) and out-of-plane ($\rho_\perp$) resistivity tensor components. The conductivity in the plane *ab* is practically isotropic ($\sigma_{\parallel TCNQ\ stacks}/\sigma_{\perp TCNQ\ stacks} \sim$ 2.0 at 300 K). Hence the obtained compound is a quasi-two-dimensional semiconductor ($\rho_\perp/\rho_\parallel \sim 1.3 \times 10^3$, inset to Fig. 6) in contrast to the majority of known TCNQ salts, which are generally the quasi-one-dimensional conductors.[4] The averaged value of the in-plane conductivity at room temperature is $5.2 \times 10^{-2}$ $\Omega^{-1}$cm$^{-1}$. With respect to the data in the Fig. 4, it shows similar exponential behavior with noticeably smaller $\Delta E$ values varying between principal axes in the plane from 0.12 to 0.15 eV. A distinct deviation from the theoretical dependence takes place at T > 270 K for the temperature curves of the in-plane resistivities measured along the principal axes (Fig. 4 and 5). Surprisingly, this feature was not observed for the out-of-plane component of resistivity, $\rho_\perp(T)$, shown in Fig. 6. This is a direct indication that the transversal transport is not affected by the spin-crossover transition.

Semiconducting behavior was expected for the trimeric dianion sublattice [(TCNQ)$_3^{2-}$]$_n$ because of overlapping $\pi$-orbitals of the TCNQ molecules within the trimer and shortened intertrimer distances 3.33 Å (Fig. 3). Although the average charge on the TCNQ$^{\delta-}$ molecules in the crystals **I** and the charge reported for two other conducting SCO complexes[11a] is the same (-0.67), the conductivity at room temperature for our crystals is an order of magnitude smaller. This difference is associated with the structure of the TCNQ stacks. In **I**, the stacks are subdivided into the pronounced triads, while in the structures described in[11a], the stacks have more regular periodicity. Peculiar feature of the crystals **I** is a quasi-two-dimensional character of conductivity due to the shortened C…N contacts between TCNQ stacks in the *ab* plane (Supporting information, Fig. S6, S7). The TCNQ molecules are not directly coordinated to the Fe(II) cations in the [Fe{HC(pz)$_3$}$_2$]$^{2+}$complexes. Therefore, it is unlikely that the *d*-orbitals of the metal ion directly interact with the $\pi$-states of the conducting electrons. However, the organic network is not completely isolated from the Fe(II) complexes in terms of cross relaxation (spin-spin interactions) due to short contacts with the ligands. This channel is of interest with regard to a manipulation of the spin polarization of the delocalized charge carriers via the switchable local moments.

The bulk static magnetic susceptibility, $\chi$, has been measured for the polycrystalline sample of **I** in the temperature range 2-400 K (Fig. 7). The two spin ensembles contribute to the susceptibility: localized moments of Fe(II) and TCNQ radicals. Below 300 K, the known complexes [Fe{HC(pz)$_3$}$_2$]$^{2+}$ with inorganic diamagnetic counterions exist in the LS state.[5c,13] As we showed in a structure characterization section the octahedral coordination site FeN$_6$ facilitates a non-magnetic ground state with the electronic configuration $t_{2g}^6 e_g^0$ and S = 0 (LS) for all the Fe(II) cations in the crystal. A reversible spin-crossover $^1A_1 \leftrightarrow ^5T_2$ usually takes place above 300 K for similar known cationic complexes.[5c,13] Hence, a weak magnetic response below 280 K, presented in terms of the product $\chi T$ (Fig. 7), is related to the TCNQ radicals. Above 280 K the $\chi T$ value dramatically increases, i.e. the [Fe{HC(pz)$_3$}$_2$]$^{2+}$ complexes undergo spin-crossover transition. A fully realized HS state was not achieved because it rests above the stability threshold for this structure (Fig. 8). The thermogravimetric analysis indicates that the structure remains stable up to 450 K (Fig. S1). We applied a symmetric Boltzmann distribution model to estimate a midpoint of the transition, T*.[15a] The $g$-factor was set as 2.0 according to the values of the effective magnetic moment in refs.[5c,15d,e,f] The magnitude boundaries for the $\chi T$ fitting curve was set between the experimental level $\chi T(100\ \text{K}) = 0.035\ \text{cm}^3\cdot\text{K}\cdot\text{mol}^{-1}$ (LS) and the estimated one $\chi T(800\ \text{K}) = 3.035\ \text{cm}^3\cdot\text{K}\cdot\text{mol}^{-1}$ (HS, grey horizontal line in Fig. 8). The boundaries take into account the paramagnetic contribution from TCNQ radicals below the transition and the theoretical estimate for S = 2 local moments plus the TCNQ response above it. A best-fit curve is shown in Fig. 8. The SCO parameters are T* = 445 K and $\Delta T$ = 200 K for the midpoint and the width of transition. The behavior of $\chi T$ at heating and cooling regimes of the measurements was reversible up to 400 K, indicating absence of significant cooperative effects. As the Fe(II) cations supposed to have a zero magnetic moment below 280 K the total magnetic response should have been attributed to the local and/or delocalized spin moments S=1/2 of the TCNQ sublattice. It was analyzed in terms of a superposition of the Curie component ($\chi_c = C/T$) and the temperature-independent paramagnetic contribution (TIP). As shown in Fig. 7 (Inset), the superposition $\chi(T) = C/T + \text{TIP}$ fits well to the experimental points. The $C$ and TIP values consisted $C = 0.034\ \text{cm}^3\cdot\text{K}\cdot\text{mol}^{-1}$ and TIP $= 0.8\cdot10^{-4}\ \text{cm}^3\cdot\text{mol}^{-1}$. Relatively small value of the constant C corresponds to the effective ~1/3 electron per a formula unit (or 1 unpaired electron for every 9 TCNQ molecules). This might be also associated with the dynamic localization of the electronic states, which changes to a thermoactivated hopping of the charge carriers in the TCNQ$^-$ sub-lattice at higher temperatures. However, what if the Curie contribution $\chi_c$ arises due to the residual Fe(II) cations in HS configuration (S=2 defects)? Field dependence of the total magnetization, $M(B)$ (Fig. 9), provides an alternative way to determine the origin of the Curie contribution at T=2 K. The Brillouin function for S = 1/2 gave a best-fit curve to the M(B) data with a scaling factor k = 0.067, which corresponded to ~7% concentration of the electronic spins (solid line in Fig. 9). Although it is somewhat smaller, this estimate correlates with the Curie contribution to the magnetic susceptibility at T<280K. A formal fitting of the $M(B)$ and $\chi T(T)$ data performed for spins S = 2, including a zero-field splitting (ZFC) parameter $D$, did not allow to get an agreement for the both sets simultaneously. For example, when a reasonable agreement was reached at $D = 8.0\ \text{cm}^{-1}$ for the $M(B)$ data, a discrepancy while fitting the $\chi T(T)$ data raised to 100%, $(\chi T)_{fit}/(\chi T)_{exp} \approx 2$ and vice versa. Moreover, a scaling factor for the best fit curve $M(B)$ was equivalent to a presence of ~3% of S=2 moments at 2 K. This does not seem reasonable taking into account the relative magnitude of changes of $\chi T$ in the range of SCO transition. Thus, a relative concentration of ~7 % seems a realistic estimate for the localized unpaired electrons in the TCNQ sublattice at helium temperatures. It is worth to note that the obtained $C$ value is an order of magnitude higher than that found in the conductors [M(dieneN$_4$)](TCNQ)$_3$ (M=Ni, Cu), whereas the TIP and resistivity values are an order of magnitude lower.[17b] The TIP values for organic conductors based on trimeric dianions TCNQ$_3^{2-}$ fit in the range of the Pauli spin susceptibilities reported for numerous conducting radical salts.[4b,15c] If so, the ratio between the TIP and the Curie contributions might correlate with the charge distribution in the trimers. Alternatively, the TIP response was also detected in some non-conducting Fe(II) LS complexes, where it was attributed to a predominately $^1A_{1g}$ state plus a contribution of a second-order Zeeman-derived temperature-independent susceptibility.[13a,b,d] So far, its origin remains unclear.

The liquid nitrogen EPR spectrum of the polycrystalline sample **I** is an intensive signal with the axial anisotropy of $g$-factor, $\Delta g = 3 \times 10^{-4}$. A central component with $g_\| = 2.005(1)$ has a Lorentzian lineshape (Supporting information, Figs. S9, S10). The lineshape of the central component was reconstructed from the high-field half of the first derivative signal. The Fig. S9 shows a comparison of the reconstructed EPR line with Lorentzian and Gaussian lines, which unambiguously indicate homogeneous mechanisms of broadening in the range of SCO (93-353 K). This speaks in favor of either motional narrowing and cross relaxation mechanisms in contrast to the supposed inhomogeneous distribution of local spin-lattice relaxation times near the HS complexes. A peak-to-peak linewidth of the total spectrum is $\Delta B_{tot}$ = 2.2 Oe, whereas for the $g_\|$ component it is $\Delta B$ = 1.6 Oe. At 283 K the entire spectrum narrows to a single symmetric line with $\Delta B(283\ \text{K})$ = 1.7 Oe (Supporting information, Fig. S10). Comparison of the partial and total linewidth evolution while heating from 93 to 285 K (Fig. 10) indicates that narrowing of the total signal is a result of isotropization of the $g$-tensor. It goes along with the decline of the anisotropy, $\rho_\perp/\rho_\|$, shown in the inset to Fig. 6. Spin-phonon interactions usually broaden the signal with the temperature.[16a] Cross relaxation and/or fast (compare to T$_2$ relaxation time) translational motion of spins in conducting samples lead to the opposite.[16b,c] However, a double integrated intensity of the EPR signal, $I_{EPR}$, does not depend on the relaxation. Its value depends on the actual spin concentration. In the interval 93-300 K $I_{EPR}$ decreases with the temperature from 2.16 × $10^{-3}$ at 93 K to 1.21 × $10^{-4}$ cm$^3$·mol$^{-1}$, which is faster than the Curie trend. Above 200 K the value of $I_{EPR}$ shows only a weak temperature dependence. As soon as the integral intensity of the absorption signal is proportional to the spin susceptibility, the product $I_{EPR}T$ was plotted in the Fig. 7 in the same scale as the $\chi T$. In the range 220-350 K the absolute value of $I_{EPR}$ and the Curie component $\chi_c$ were close within 20%. The above characteristics reliably attribute the EPR signal to the conducting electrons of the TCNQ sublattice. The signal from the Fe(II) local moments of HS Fe(II) has not been detected. This is understandable in terms of relatively low concentration (~4% at 360 K) and fast relaxation. In turn, a high relaxation rate in the S=2 spin reservoir provides the affective cross relaxation channel for the unpaired electron spins in the conducting layers.

Above 280 K the width of the TCNQ signal increased exponentially (Fig. 10). This correlates with the appearance of the local magnetic moments in the cation sublattice as well as with the structural defects (distorted TCNQ trimers) induced by the changed geometry of the individual HS Fe(II) complexes. Due to the thermal activation mechanism the concentrations of the S=2 moments as well as the defects follow similar exponential trend. However, the magnitude factors are different: the defect concentration reaches its maximum at $T^*$ = 445 K, while the concentration of the local moments saturates at ($T^*+\Delta T/2$) = 545 K. As soon as stability threshold was close to the $T^*$ value we were not able to distinguish the contributions from the above sources. For various low-dimensional organic conductors the value of the EPR linewidth usually weakly depends on the crystal quality and to a certain degree it also weakly depends on the external disorder.[15c] For a random spatial distribution of the defects one could rather expect the inhomogeneous broadening mechanism than a systematic change of the spin-lattice relaxation rate due to continuous evolution of the electronic spectrum of the TCNQ structure. However, the lineshape remains Lorentzian within the studied SCO range. The inset of Fig. 10 demonstrate the evolution of the linewidth versus the concentration of the local moments S=2 in the cation sublattice. The value of the relative concentration, $n/N$(%), was extracted from the magnetic susceptibility data on Fig. 7. In the SCO range the linewidth was proportional to the squared concentration, $\Delta B=\Delta B_0(1+k(n/N)^2)$, where $\Delta B_0$=1.7 Oe and $k$= 1.15·$10^{-1}$ Oe. Surprisingly, the resistivity reveals a weaker sensitivity to the SCO. A value of relative deviation for the *a*- and *b*- components of the in-plane resistivity, $\Delta_\rho(\%) = (\rho_\parallel(T) - \rho_{\parallel calc}(T))/\rho_{\parallel calc}(T)$, shown in Fig. 11, was found to reveal a universal logarithmic dependence on the concentration $n/N$, $\Delta_\rho(\%)=A+B·ln(n/N)$, where $A$=0.60 and $B$=8.51·$10^{-2}$. Here the $\rho_{\parallel calc}(T)$ values were extracted from the *a*- and *b*- best fit curves extended to the higher temperatures (dashed lines in Fig. 5). The transverse transport, $\rho_\perp(T)$, did not react to the changes in the cation layers and therefore did not contribute to the line broadening effect. Qualitatively this means that the probability of Zeeman spin transitions grows faster than the in-plane scattering rate (or in-plane hopping rate) of the charge carriers, i.e. phenomenological Elliott or Weger approaches are not applicable.[17] This discrepancy could be explained by the activation of additional spin relaxation channel related to the expanding reservoir of local moments in [Fe{HC(pz)$_3$}$_2$]$^{2+}$ complexes. A cross relaxation due to spin-spin interactions between the two spin reservoirs: local moments of Fe(II) ions in HS state and delocalized spin moments of the conduction electrons, is likely to contribute to the EPR line broadening in the SCO range but do not affect the resistivity. Even weak interactions via short contacts are able to provide an efficient relaxation channel (reciprocal bottleneck effect).[18] At present one cannot determine whether the dipole-dipole or weak superexchange interactions facilitate the cross relaxation in [Fe{HC(pz)$_3$}$_2$](TCNQ)$_3$. The important consequence is that the local magnetic moments in the Fe(II) complexes not just coexist but affect the spin degrees of freedom in the conducting TCNQ layers. In a sense, the SCO transition triggers a synergy of the two subsystems.

**CONCLUSION**

We prepared and structurally characterized the conducting spin-crossover compound based on Fe(II) cation complex with the electroactive trimeric dianionic TCNQ-network in a fractional reduction state, [Fe{HC(pz)$_3$}$_2$](TCNQ)$_3$ (**I**). In the crystals of **I**, there are two crystallographic independent TCNQ species in the different charge states (labeled as TCNQ (**A**) and TCNQ (**B**)) in an asymmetric unit. The unit A and B stack in the pattern …ABA…ABA… with interplanar distances of 3.03 Å and 3.33 Å inside triads and between triads, respectively. The conductivity along the stacks is 0.015 Ω$^{-1}$cm$^{-1}$ at room temperature. The shortened C…N contacts between the stacks gives rise to a quasi-two-dimensional character of conductivity in the ab plane. The compound **I** is only the second example of the conducting SCO systems that combine Fe(II) SCO cation complexes with partially charged TCNQ$^{\delta-}$ radical anions in the same crystal lattice. It was found that a robust broadening of the TCNQ EPR signal and a pronounced deviation of the in-plane resistivity of the conducting layers out of the exponential law are induced by the reversible SCO transition in the [Fe{HC(pz)$_3$}$_2$]$^{2+}$ complexes. The enhancement of spin depolarization of the charge carriers triggered be the SCO transition was interpreted in terms of cross relaxation between the two spin reservoirs: local moments of Fe(II) ions and delocalized spin moments of the conduction electrons. Reasoning from the synthetic strategy used in the work, one may expect that a variety of conducting compounds comprising Fe(II) SCO complexes may be significantly expanded.

## EXPERIMENTAL

### Synthesis of [Fe{HC(pz)$_3$}$_2$](TCNQ)$_3$ (I)

The crystals of **I** were obtained in argon atmosphere by mixing the boiling solutions [Fe{HC(pz)$_3$}$_2$]I$_2$ H$_2$O (0.13 mmol) in 30 ml of MeOH/abs.EtOH/MeCN (1:1:1) and TCNQ (0.26 mmol) in MeCN (10 ml) followed by slow cooling of the resulting solution. After two days, black shining crystals in the shape of hexagons were formed. The crystals were collected by filtration, washed with ethanol, ether, and dried under vacuum at room temperature. Yield: 75%. Anal. Calcd. (Found) for **I** (C$_{56}$H$_{32}$N$_{24}$Fe): C, 61.32 (61.38); H, 2.94 (3.15); N, 30.65 (30.30) %. It is worth to note that the same crystals were formed at the ratios of starting reagents of 1:3 and 1:4. It should be note that a hexagon shape of the obtained crystals (Supporting information, Figs. S8) is not typical for TCNQ salts, which generally have a needle-like shape.

### Thermogravimetric analysis

The thermogravimetric analysis was performed in argon atmosphere with a heating rate 5.0 °C min$^{-1}$ using a NETZSCH STA 409 C Luxx thermal analyzer, interfaced to a QMS 403 Aelos mass spectrometer, which allows simultaneous thermogravimetry (TG), differential scanning calorimetry (DSC) and mass-spectrometry measurements.

### X-ray Crystallography

Crystals of (**I**) (C$_{20}$H$_{20}$FeN$_{12}$·3C$_{12}$H$_4$N$_4$, FW = 1096.91) are triclinic, space group P-1, at 100K: $a$ = 8.1253(13), $b$ = 10.0214(16), $c$ = 16.747(3) Å, $\alpha$ = 100.977(3), $\beta$ = 91.048(3), $\gamma$ = 111.167(3)°. V = 1242.7(3) Å$^3$, Z = 1 (Z' = 0.5), d$_{calc}$ = 1.466 gcm$^{-3}$, μ(MoKα) = 3.72 cm$^{-1}$, F(000) = 562. Intensities of 11053 reflections were measured with a Bruker SMART APEX2 CCD diffractometer [λ(MoKα) = 0.71072Å, ω-scans, 2θ < 58°] and 6509 independent reflections [R$_{int}$ = 0.0291] were used in further refinement. The structure was solved by direct method and refined by the full-matrix least-squares technique against F$^2$ in the anisotropic-isotropic approximation. Hydrogen atoms were located from the Fourier synthesis of the electron density and refined in the isotropic approximation. For (**I**) the refinement converged to wR2 = 0.1231 and GOF = 1.027 for all independent reflections (R1 = 0.0452 was calculated against F for 4941 observed reflections with I > 2σ(I))**.** All calculations were performed using SHELXTL PLUS 5.0.

CCDC 1033050 contains supplementary crystallographic data for this paper. These data can be obtained free of charge from the Cambridge Crystallographic Data Centre via www.ccdc.cam.ac.uk/data_request/cif.

### Transport and magnetic measurements

The dc resistivity measurements were performed on single crystals by a standard four-probe method with the current flow parallel to the TCNQ stacks in the temperature range 120-360 K. Four annealed platinum wires (0.02 mm in diameter) were attached to a crystal surface by a graphite paste (Supporting information, Fig. S8). The applied current was in the limits 10-100 nA. This geometry is convenient for the test measurements to reveal the features in the temperature dependences of the resistance, but in the strongly anisotropic sample the measured value includes the mixture of both in-plane and out-of-plane components of the resistivity tensor, because the current is distributed non-uniformly through the sample cross section. This is why in the control experiments we have measured the resistivity tensor components separately. To measure in-plane anisotropy we have used Montgomery method[19] on the samples, which had the shape of the thin hexagon plates elongated in the direction of TCNQ stacks (the typical sample shape is shown in the Supporting information, Fig. S8). So, using two pairs of contacts attached to the plate corners on the long sides of the plate we could measure two components of the resistivity tensor along and perpendicular to the of TCNQ stacks. To measure the out-of-plane resistivity tensor we have used the modified Montgomery method[20] on the sample with two pairs of contacts attached to the opposite sample surfaces.

Magnetic measurements were carried out by using a Quantum Design MPMS-5-XL SQUID magnetometer. The static magnetic susceptibility $\chi(T)$ of the polycrystalline sample was measured at the magnetic field $B = 0.1$ T, at cooling and heating regimes in the temperature range of 2-400 K. Field dependences of the magnetization $M(B)$ were obtained at 2.0 K in the field range of -5.0 to +5.0 T. The sample had been cooled down to 2.0 K in a weak magnetic field $B = 0.1$ T. Then the measurements were performed at increasing field up to 5.0 T and further field decreasing with a sign reversal down to -5.0 T.

EPR spectra were recorded in the temperature range of 90-350 K on a standard homodyne X-band Bruker Elexsys E580 FT/CW X-Band spectrometer (9.4 GHz). The temperature was set and stabilized at a rate of 1-2 K min$^{-1}$ with an accuracy of 0.1 K using a liquid nitrogen gas-flow cryostat. The spin contribution to the magnetic susceptibility was determined by the double integration of the EPR signal (Schumacher-Slichter method) under conditions for the field sweep $\delta B_{sw} \geq 10\Delta B$ ($\Delta B$ is the peak-to-peak EPR linewidth). In this case, the error of the method for the Lorentz-type EPR signal is ~10%. The pyrolytic coal product with $g = 2.00283$ was used as the standard of a spin concentration.

## ASSOCIATED CONTENT

**Supporting information**

Thermogravimetric data, various structure figures and tables, additional EPR data, photo of a crystal with contacts for measurement of conductivity, crystallographic CIF file CCDC 1033050.

## AUTHOR INFORMATION

**Corresponding Authors**

*E-mail: starichenko@imp.uran.ru

*E-mail: yuri.shvachko@gmail.com

*E-mail: yagubski@icp.ac.ru

## ACKNOWLEDGMENT

This work was supported by the Russian Foundation for Basic Research, projects No. 14-03-00119, 15-02-02723 and by grant for support of the leading scientific schools No. 1540.2014.2. The authors would like to thank P. Barzilovich for his assistance in X-ray experiment and L. Zorina for enlightening discussion of X-ray results.

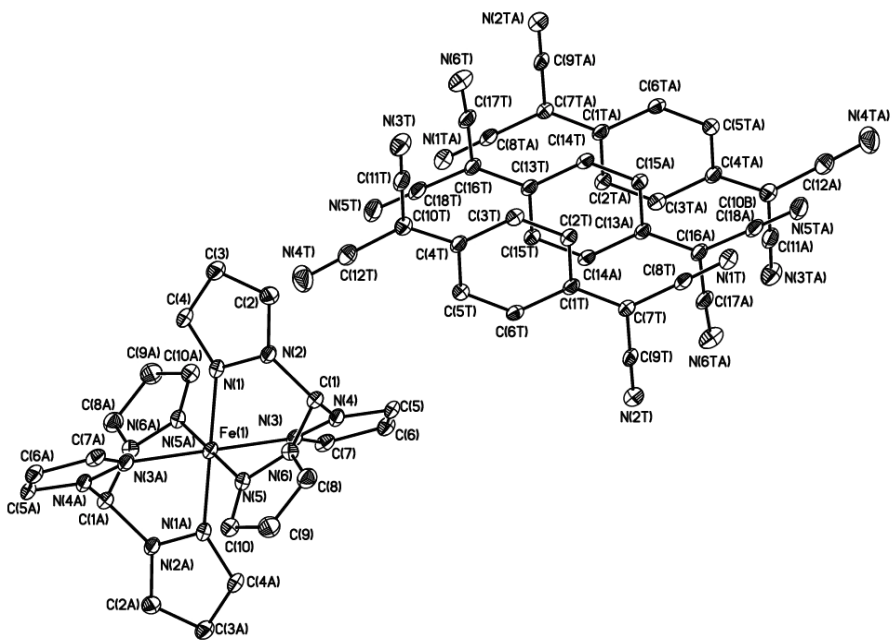

**Fig. 1.** A formula unit of complex **I**. The $\{Fe[HC(pz)_3]_2\}^{2+}$ cation and one of the TCNQ molecules occupy the special position (the center of symmetry). All atoms with label **A** are obtained from the base ones by the symmetry operations (1 -x, -y, -z for cation and 2-x+1, -y+2, -z+1 for TCNQ).

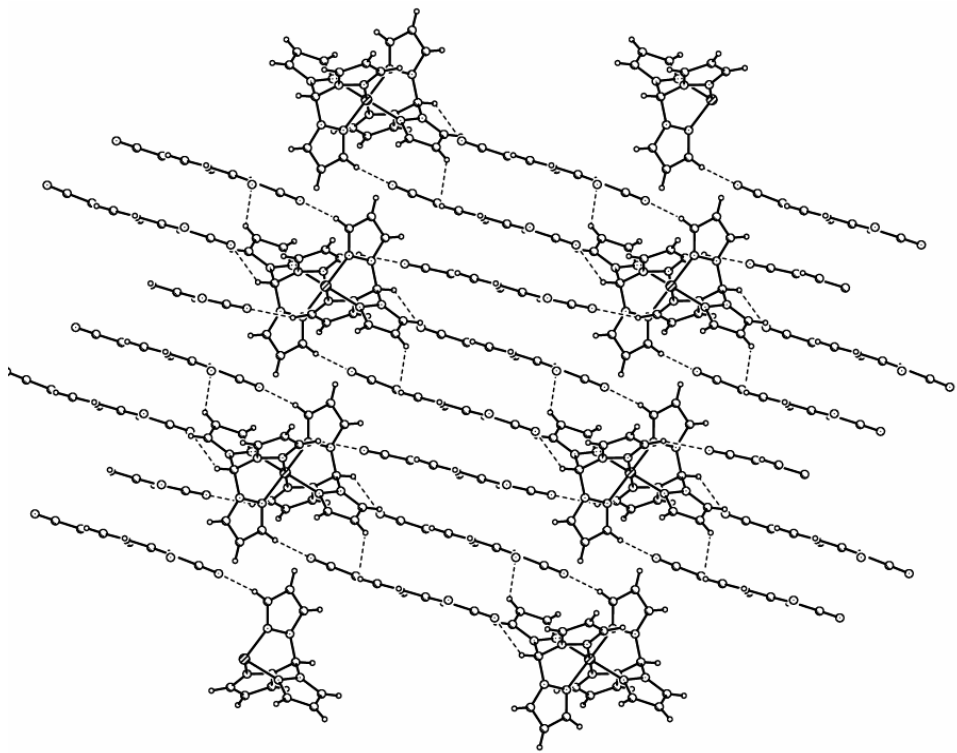

**Fig. 2.** Fragment of the crystal packing in **I** snowing alternate cationic and anionic layers (view along *a*). Interlayer C-H…N contacts (2.41-2.70 Å) between cations and anions are shown by dashed lines.

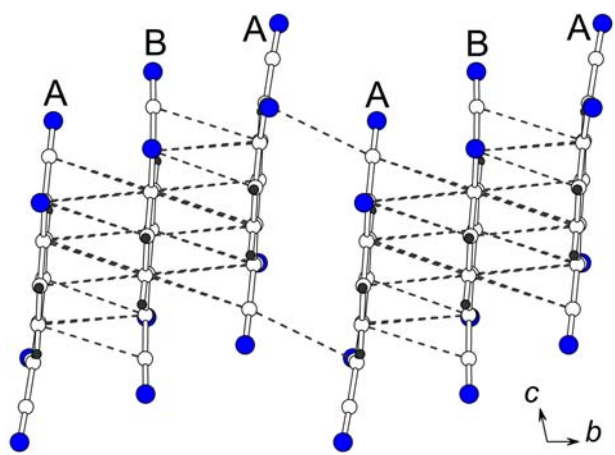

**Fig. 3.** Stacking interactions in the anionic columns.

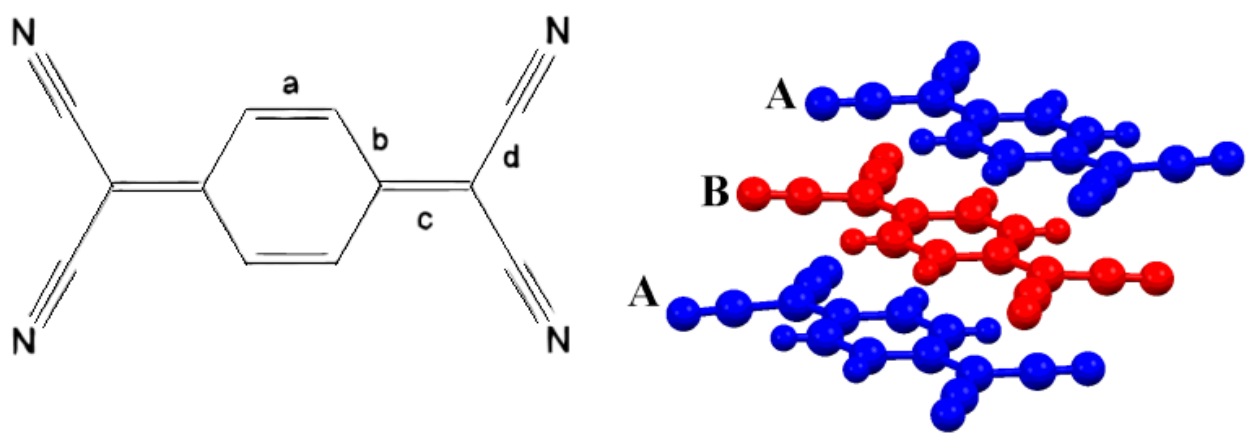

| Compounds | a | b | c | d | c/(b+d) | δ |
|---|---|---|---|---|---|---|
| TCNQ$^0$ | 1.346 | 1.448 | 1.374 | 1.440 | 0.476 | 0.0[14b] |
| TCNQ$^{-1}$ | 1.373 | 1.423 | 1.420 | 1.416 | 0.500 | -1.0[14b] |
| TCNQ$^{-0.5}$ | 1.354 | 1.434 | 1.396 | 1.428 | 0.488 | -0.5[14b] |
| TCNQ (**I**, **A**) | 1.359 | 1.437 | 1.399 | 1.428 | 0.488 | -0.5 |
| TCNQ (**I**, **B**) | 1.371 | 1.430 | 1.420 | 1.425 | 0.497 | -0.89 |

**Table 1.** The charges (δ) of different TCNQ species in [Fe{HC(pz)$_3$}$_2$](TCNQ)$_3$ (**I**) estimated from Kistenmacher's empirical formula.

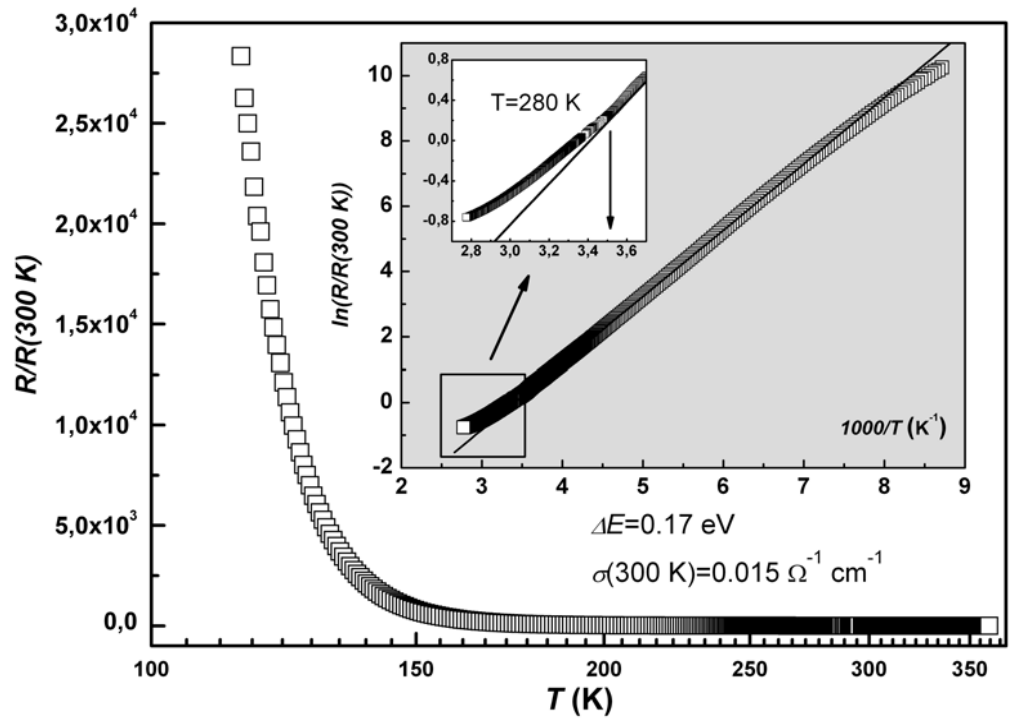

**Fig. 4.** Temperature dependence of the normalized resistance, $R(T)/R(300\ \text{K})$. Inset: logarithmic plot of the $R(T)/R(300\ \text{K})$ vs. scaled reciprocal temperature, $1000/T$. Solid line, $R_{calc}(T) = \exp(\Delta E/kT)$ with the energy gap $\Delta E = 0.17$ eV, is the best-fit ($R_{factor} = 0.99977$) of the experimental data in the range $120 < T < 250$ K. A deviation of the experimental data $R(T)$ from the theoretical curve $R_{calc}(T)$ at $T > 280$ K is shown in details in the zoomed window.

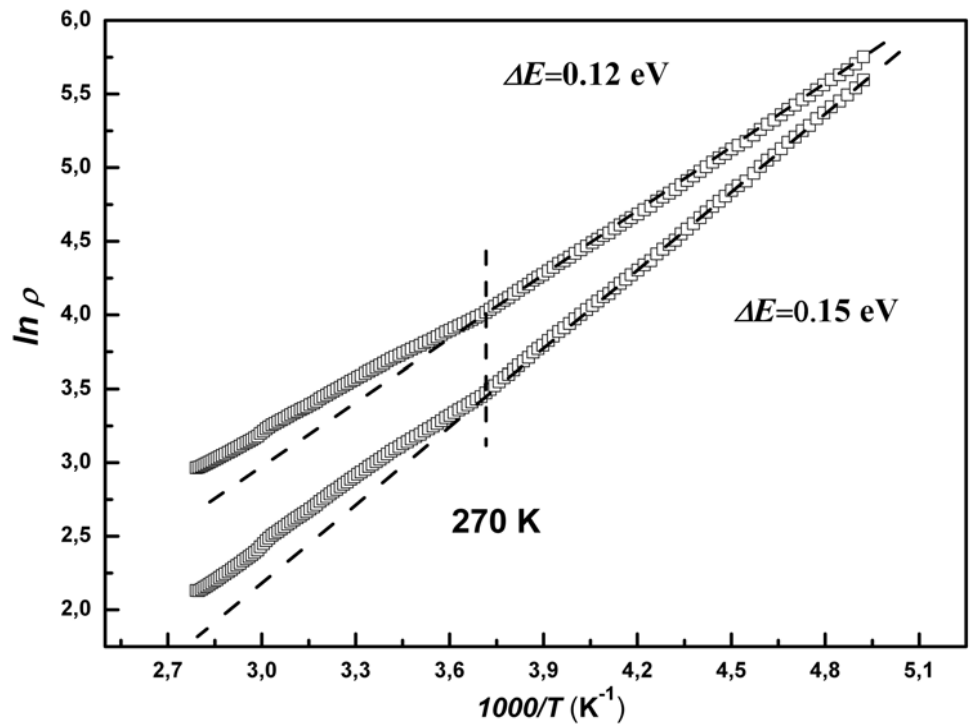

**Fig. 5.** Temperature dependence of the in-plane (*ab*) resistivities measured by Montgomery method. The vertical dashed line corresponds to the temperature at which the deviation from the exponential dependence takes place.

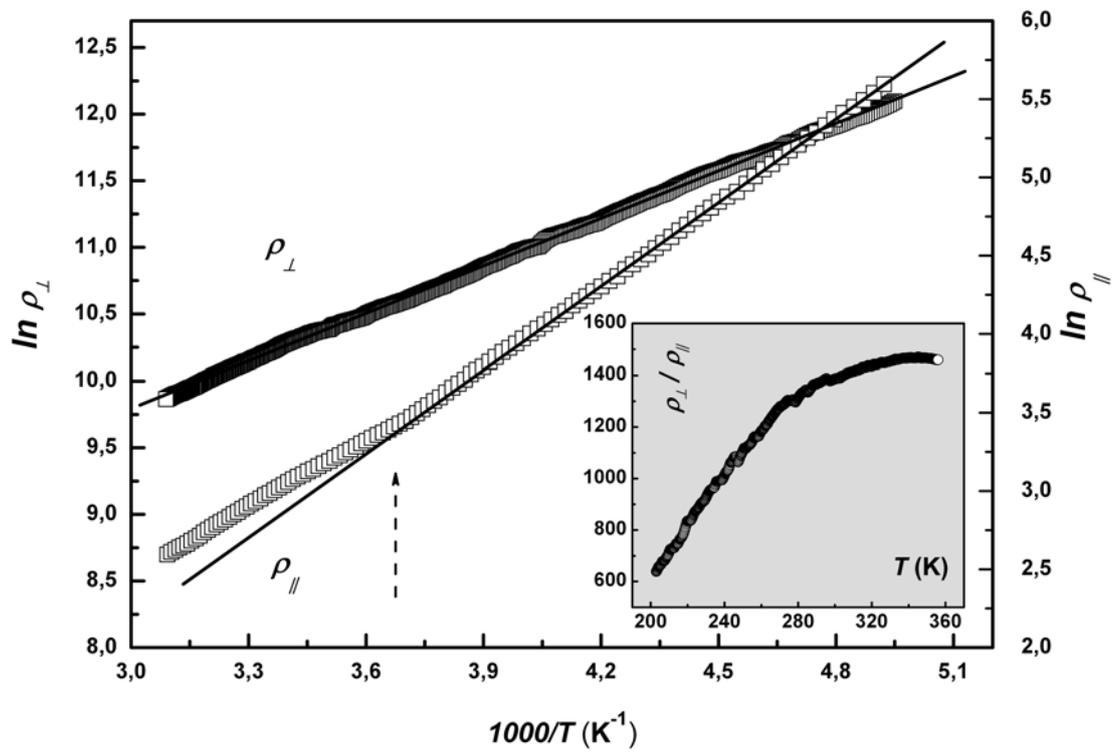

**Fig. 6.** Temperature dependence of the in-plane and out-of-plane resistivities measured by Montgomery method. Inset: temperature dependences of the resistivity anisotropy $\rho_\perp/\rho_\parallel$.

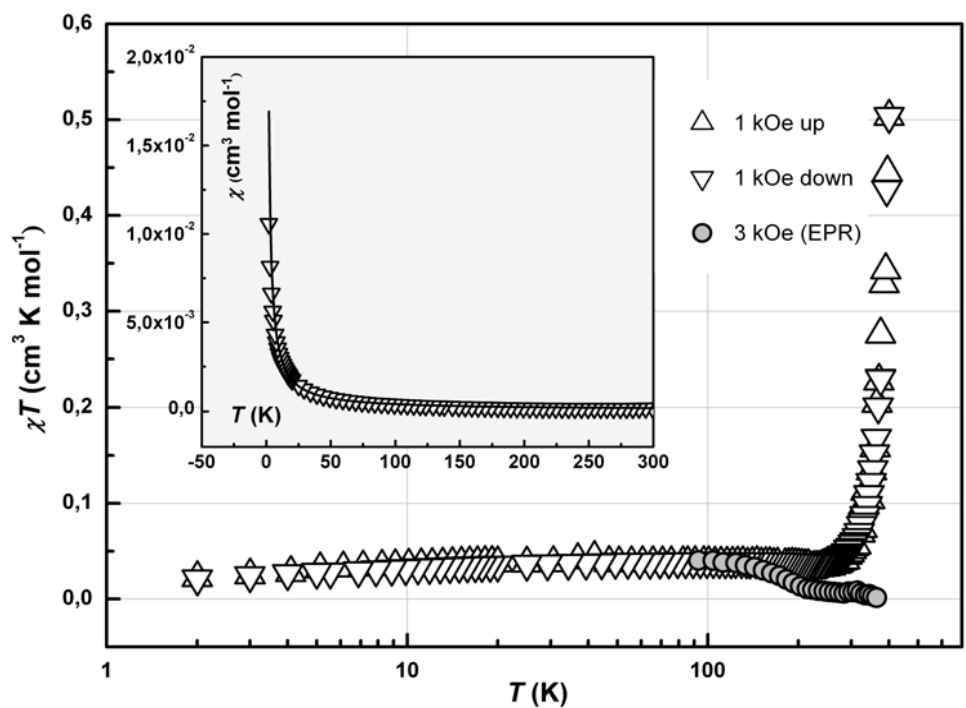

**Fig. 7.** Temperature dependences of the product $\chi T$ ($B = 1$ kOe, cooling ($\triangledown$) and heating ($\triangle$) regimes, SQUID). Filled circles – the product $I_{EPR}T$ measured at $B = 3$ kOe by EPR. Inset: temperature dependences of the static magnetic susceptibility $\chi(T)$. The solid line is a best fit to a Curie law with $C = 0.034$ cm$^3$·K·mol$^{-1}$ plus a TIP contribution of $0.80 \cdot 10^{-4}$ cm$^3$·mol$^{-1}$.

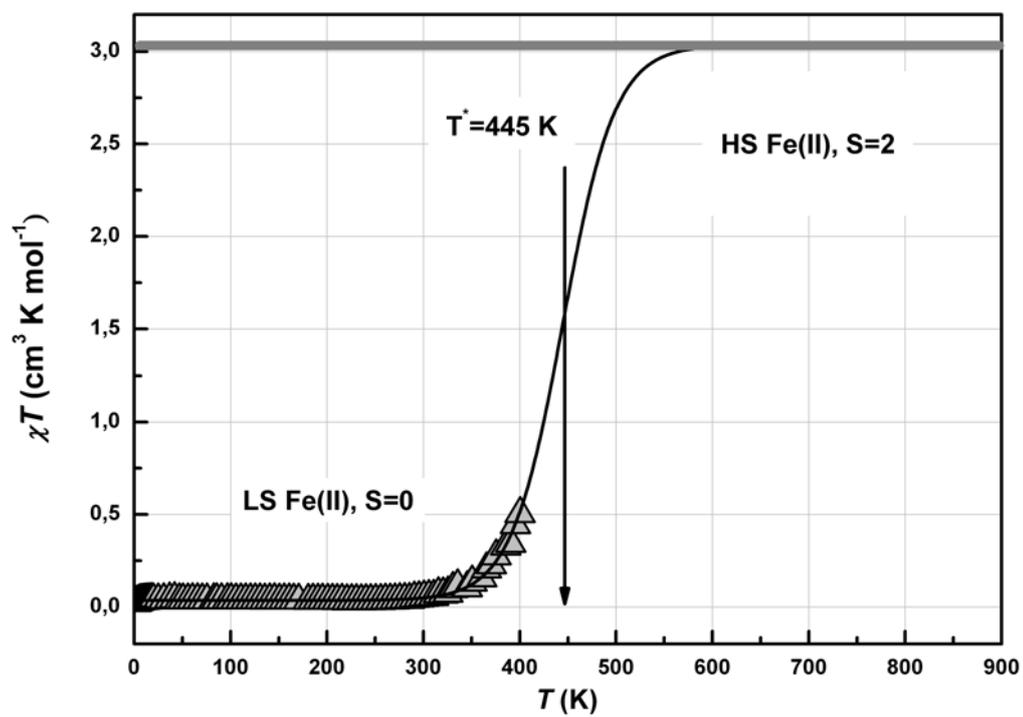

**Fig. 8.** Spin-crossover transition in [Fe{HC(pz)$_3$}$_2$](TCNQ)$_3$. Triangles are the experimental $\chi T$ data. Solid line is simulation by a Boltzmann distribution, $\chi T$(100 K) = 0.035 cm$^3$·K·mol$^{-1}$, $\chi T$(800 K) = 3.035 cm$^3$·K·mol$^{-1}$.

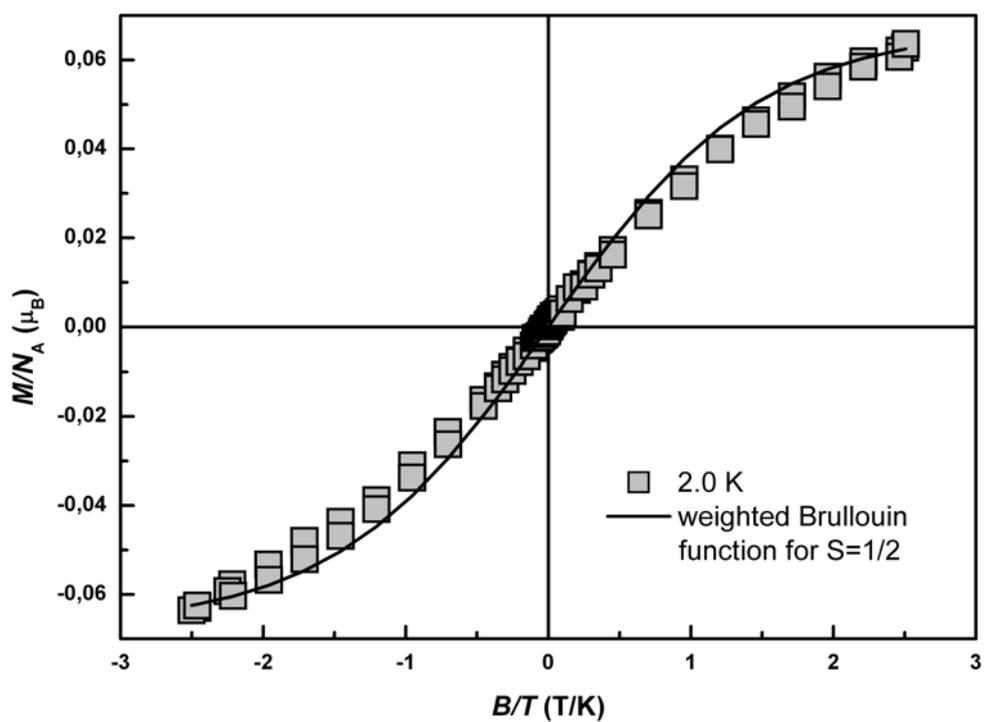

**Fig. 9.** Field dependence of the magnetization, $M(B)$, measured at T = 2.0 K. Solid line is Brillouin function approximation with S = 1/2 and scaling factor k = 0.067.

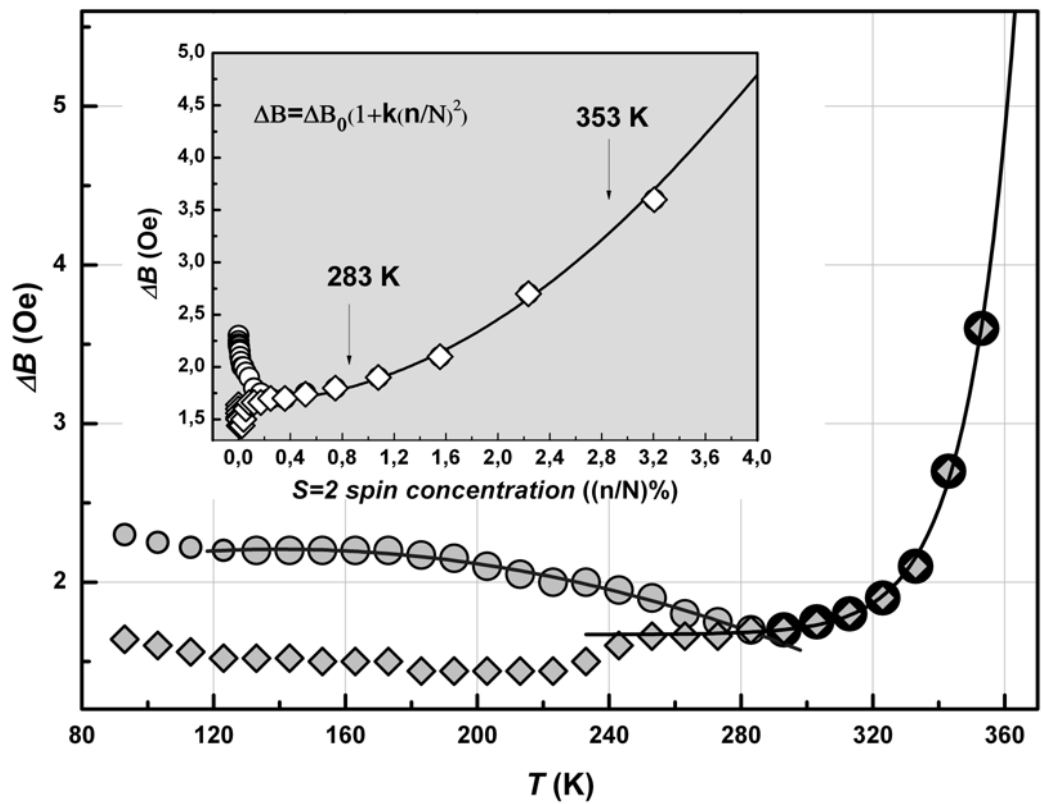

**Fig. 10.** Temperature dependence of the EPR linewidth, $\Delta B(T)$: circles – peak-to-peak linewidth of the total EPR spectrum; diamonds – peak-to-peak linewidth of the $g_{\parallel}$ component. Solid line for grey circles is a polynomial fit (narrowing), for black circles – exponential fit (broadening). Inset: EPR linewidth, $\Delta B(T)$, vs. concentration of the local moments S=2 in the cation sublattice, $n/N(\%)$.

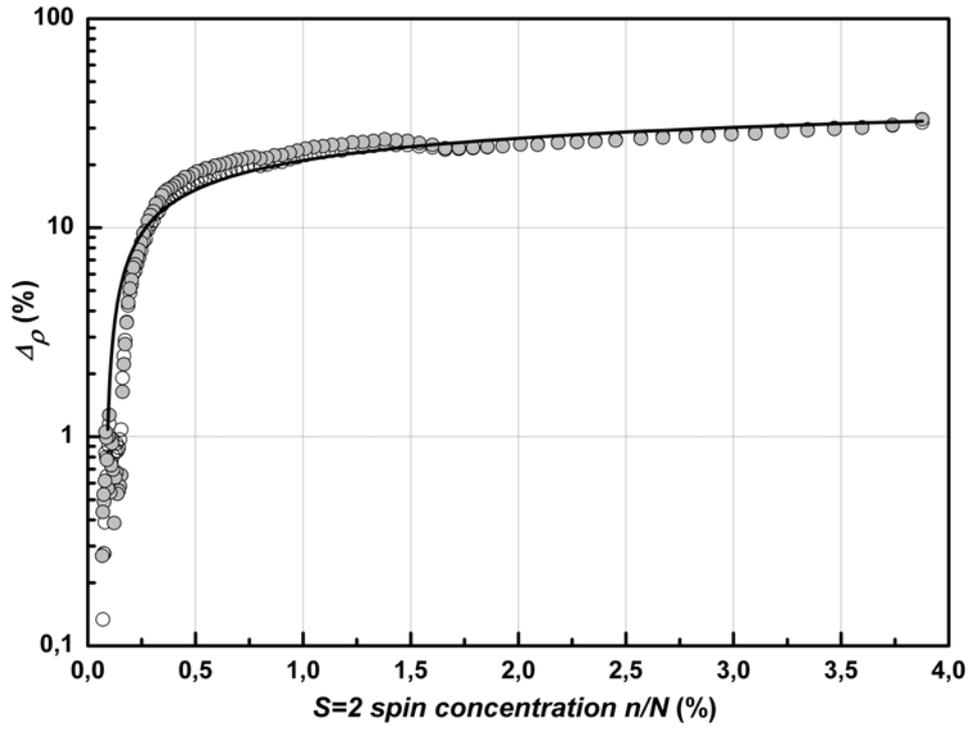

**Fig. 11.** A relative deviation for the *a*- and *b*- components of the in-plane resistivity, $\Delta_\rho(\%) = (\rho_\parallel(T) - \rho_{\parallel calc}(T))/\rho_{\parallel calc}(T)$, versus the concentration of S=2 local moments in the range of the SCO transition. The $\rho_{\parallel calc}(T)$ values were obtained from the *a*- and *b*- best fit curves extended to the higher temperatures (dashed lines in Fig. 5). A solid line is the best fit logarithmic curve, $\Delta_\rho(\%)=A+B\cdot ln(n/N)$, where $A=0.60$ and $B=8.51\cdot 10^{-2}$. Here the *n/N(%)* values were extracted from the modelling curve in Fig. 8.

# Supporting Information

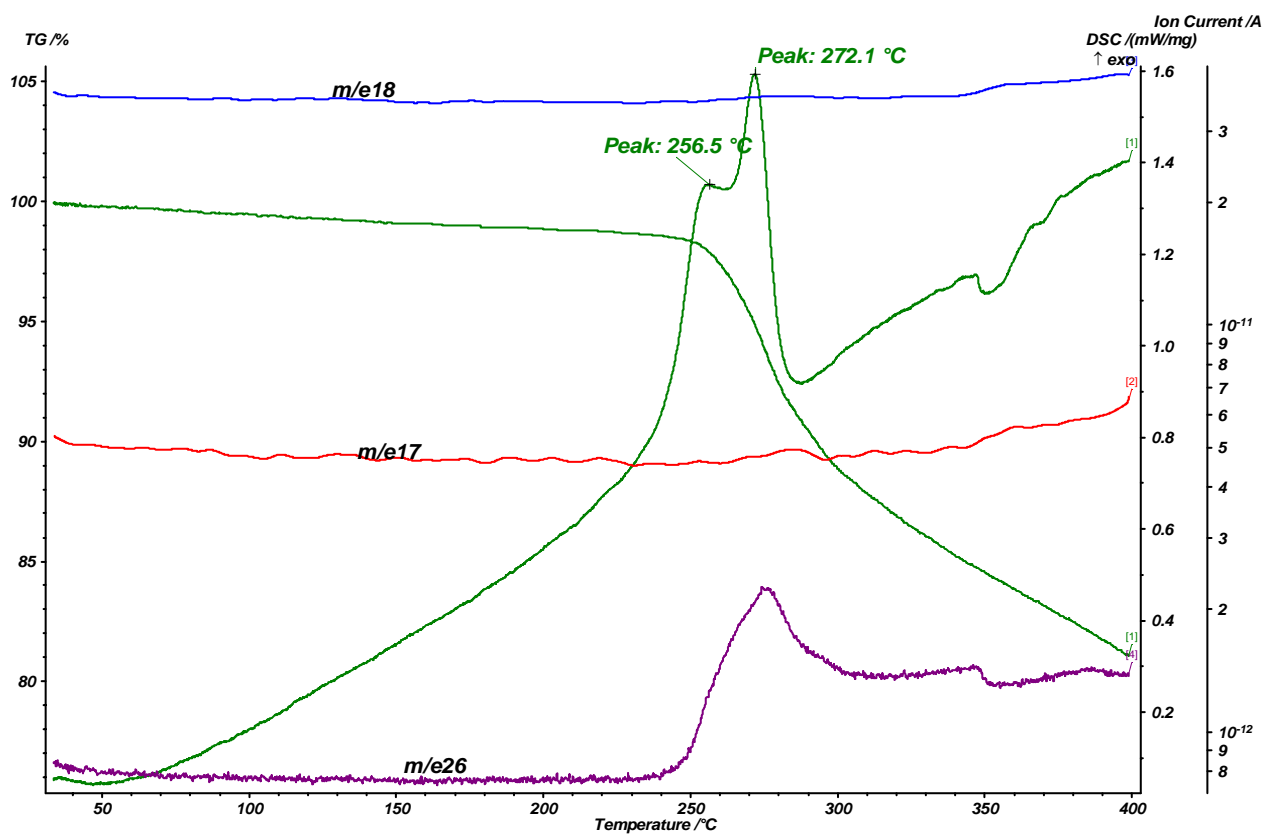

**Fig. S1.** TG-DSC curves and mass spectra for complex **I**.

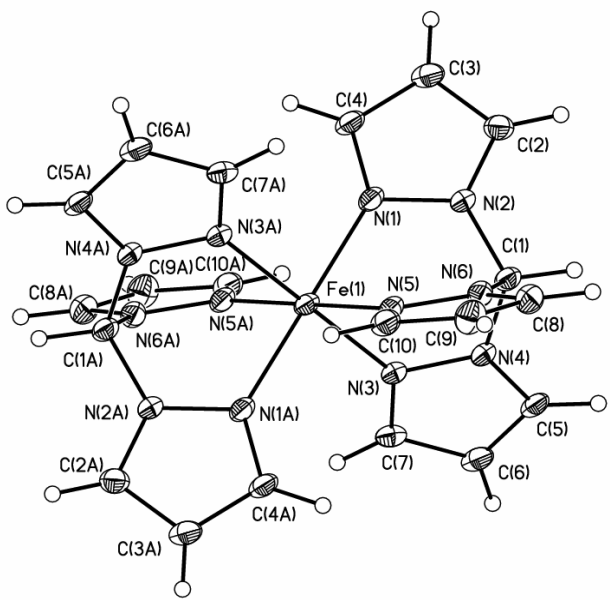

**Fig. S2** General view of the cation in **I** in representation of atoms by thermal ellipsoid plots (p=50%).

**Table S1.** Bond lengths (Å) and angles in [Fe{HC(pz)₃}₂]²⁺ (°).

| Bond lengths (Å) | | | |
|---|---|---|---|
| Fe(1)-N(1) | 1.9644(16) | N(5)-C(10) | 1.333(3) |
| Fe(1)-N(5) | 1.9672(17) | N(5)-N(6) | 1.369(2) |
| Fe(1)-N(3) | 1.9735(16) | N(6)-C(8) | 1.352(3) |
| N(1)-C(4) | 1.335(2) | N(6)-C(1) | 1.445(2) |
| N(1)-N(2) | 1.367(2) | C(2)-C(3) | 1.379(3) |
| N(2)-C(2) | 1.356(2) | C(3)-C(4) | 1.409(3) |
| N(2)-C(1) | 1.453(2) | C(5)-C(6) | 1.363(3) |
| N(3)-C(7) | 1.323(2) | C(6)-C(7) | 1.407(3) |
| N(3)-N(4) | 1.369(2) | C(8)-C(9) | 1.376(3) |
| N(4)-C(5) | 1.363(2) | C(9)-C(10) | 1.402(3) |
| N(4)-C(1) | 1.444(2) | | |
| Bond angles (°) | | | |
| N(1)-Fe(1)-N(1A)[a)] | 80 | N(4)-N(3)-Fe(1) | 118.66(12) |
| N(1)-Fe(1)-N(5A) | 91.88(7) | C(5)-N(4)-N(3) | 111.70(16) |
| N(1A)-Fe(1)-N(5A) | 88.12(7) | C(5)-N(4)-C(1) | 129.90(16) |
| N(1)-Fe(1)-N(5) | 88.12(7) | N(3)-N(4)-C(1) | 117.74(14) |
| N(1A)-Fe(1)-N(5) | 91.88(7) | C(10)-N(5)-N(6) | 105.40(16) |
| N(5A)-Fe(1)-N(5) | 180 | C(10)-N(5)-Fe(1) | 136.16(14) |
| N(1)-Fe(1)-N(3A) | 94.47(6) | N(6)-N(5)-Fe(1) | 118.44(12) |
| N(1A)-Fe(1)-N(3A) | 85.53(6) | C(8)-N(6)-N(5) | 111.55(16) |
| N(5A)-Fe(1)-N(3A) | 88.41(7) | C(8)-N(6)-C(1) | 130.12(16) |
| N(5)-Fe(1)-N(3)#1 | 91.59(7) | N(5)-N(6)-C(1) | 118.30(16) |
| N(1)-Fe(1)-N(3) | 85.53(6) | N(4)-C(1)-N(6) | 109.72(15) |
| N(1A)-Fe(1)-N(3) | 94.47(6) | N(4)-C(1)-N(2) | 108.65(16) |
| N(5A)-Fe(1)-N(3) | 91.59(7) | N(6)-C(1)-N(2) | 109.45(15) |
| N(5)-Fe(1)-N(3) | 88.41(7) | N(2)-C(2)-C(3) | 106.55(17) |
| N(3A)-Fe(1)-N(3) | 180 | C(2)-C(3)-C(4) | 105.62(17) |
| C(4)-N(1)-N(2) | 105.16(15) | N(1)-C(4)-C(3) | 110.78(17) |
| C(4)-N(1)-Fe(1) | 136.61(14) | N(4)-C(5)-C(6) | 106.47(17) |
| N(2)-N(1)-Fe(1) | 117.97(11) | C(5)-C(6)-C(7) | 105.72(17) |
| C(2)-N(2)-N(1) | 111.89(15) | N(3)-C(7)-C(6) | 111.43(18) |
| C(2)-N(2)-C(1) | 129.02(16) | N(6)-C(8)-C(9) | 106.52(18) |
| N(1)-N(2)-C(1) | 118.92(15) | C(8)-C(9)-C(10) | 106.10(19) |
| C(7)-N(3)-N(4) | 104.66(15) | N(5)-C(10)-C(9) | 110.43(17) |
| C(7)-N(3)-Fe(1) | 136.65(14) | | |
| Torsion angles (°) | | | |
| Fe(1)-N(1)-N(2)-C(2) | 174.77(13) | Fe(1)-N(3)-N(4)-C(5) | -178.12(12) |
| Fe(1)-N(5)-N(6)-C(8) | 180.00(13) | | |

a) All atoms with label **A** are obtained from the base ones by the symmetry operations 1-x,-y,-z.

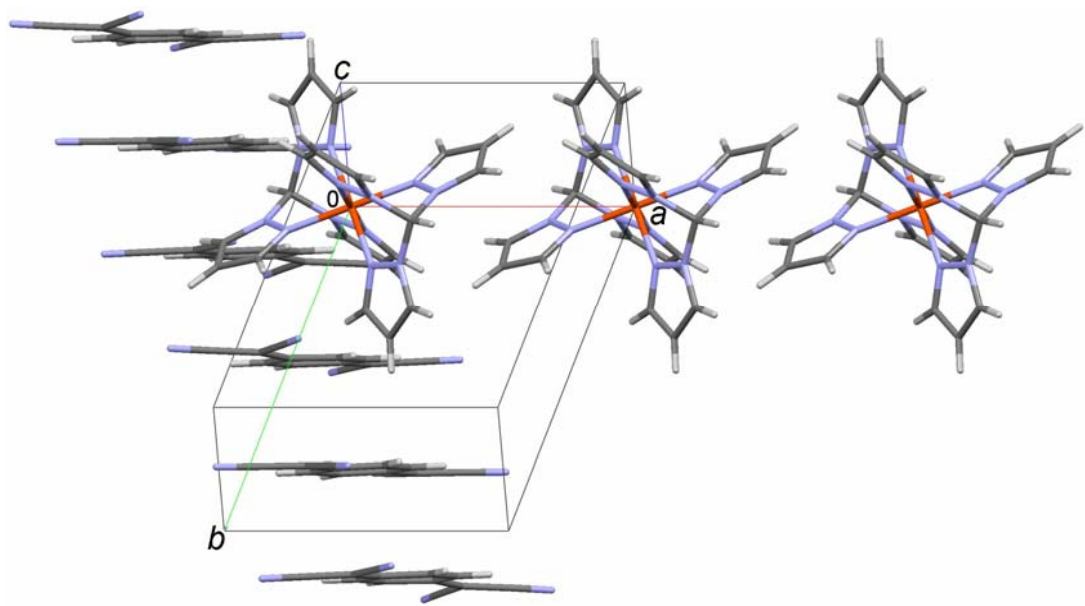

**Fig. S3.** The projection of the structure **I** on a plane *ab*.

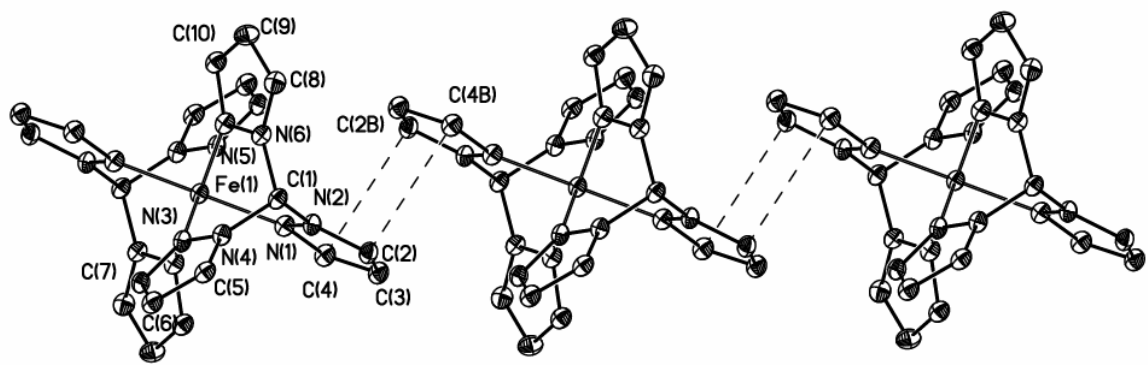

**Fig. S4.** Stacking-bonded cationic chain in the crystal of **I**.

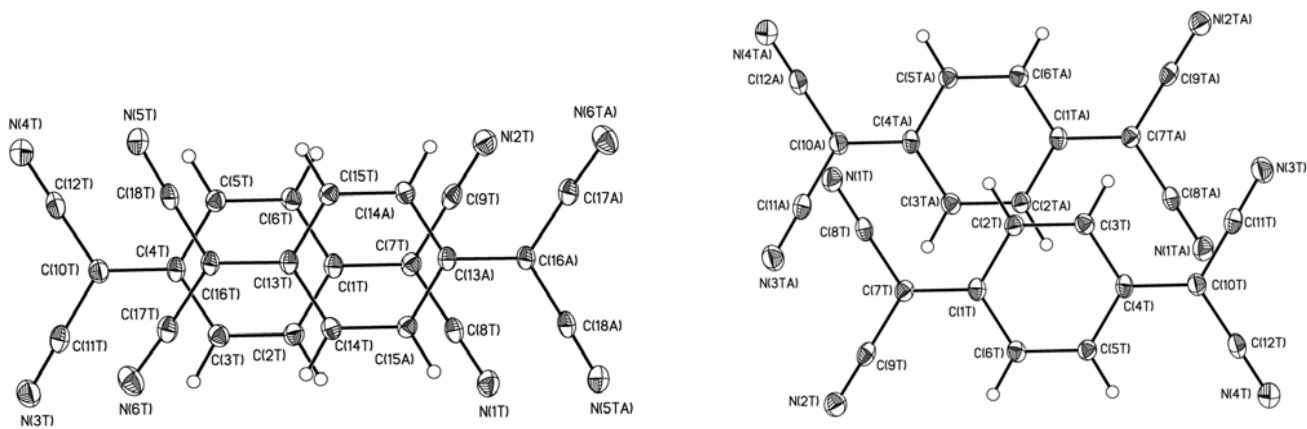

**(a)** **(b)**

**Fig. S5.** The character of TCNQ overlap within the **ABA** triads (a) and between triads (b).

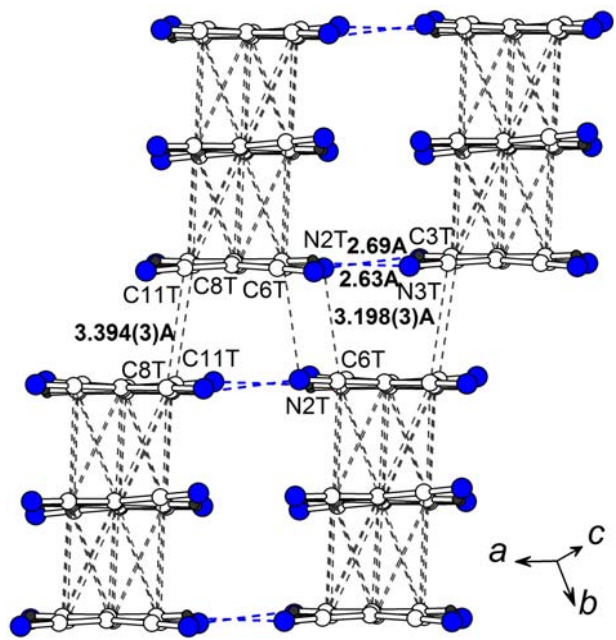

**Fig. S6.** View of the TCNQ layer along the long molecular axis of TCNQ. The TCNQ stacks are parallel to *ab* diagonal. The TCNQ triads interact by C…C intrastack contacts along [110] and C…N interstack contacts along [010] as well as weak side-by-side C-H…N hydrogen bonds.

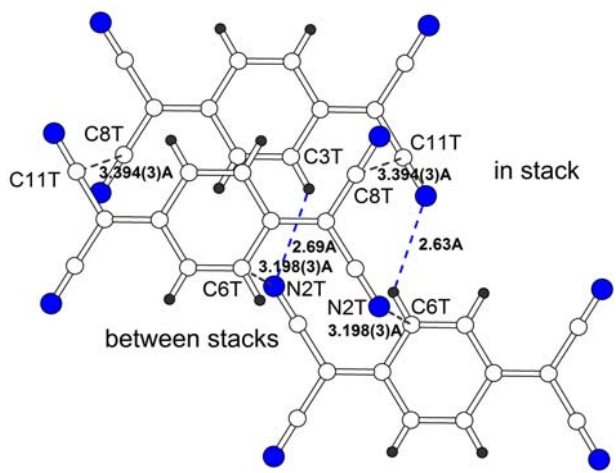

**Fig. S7.** Overlap mode and intermolecular interactions between the TCNQ triads in the stack (upper two molecules) and between the stacks (lower two molecules).

**Table S2.** Bond lengths (Å) and angles (°) in TCNQ molecules (**A** and **B**).

| A | | B | |
|---|---|---|---|
| N(1T)-C(8T) | 1.152(3) | N(5T)-C(18T) | 1.158(3) |
| C(1T)-C(7T) | 1.399(2) | N(6T)-C(17T) | 1.154(3) |
| C(1T)-C(6T) | 1.434(3) | C(13T)-C(16T) | 1.420(2) |
| C(1T)-C(2T) | 1.440(3) | C(13T)-C(14T) | 1.426(3) |
| N(2T)-C(9T) | 1.151(3) | C(13T)-C(15T) | 1.435(3) |
| C(2T)-C(3T) | 1.356(2) | C(14T)-C(15TA)[a] | 1.371(2) |
| C(3T)-C(4T) | 1.437(3) | C(15T)-C(14TA) | 1.371(2) |
| N(3T)-C(11T) | 1.151(3) | C(16T)-C(18TA) | 1.421(3) |
| N(4T)-C(12T) | 1.148(3) | C(16T)-C(17TA) | 1.430(3) |
| C(4T)-C(10T) | 1.399(2) | | |
| C(4T)-C(5T) | 1.438(3) | | |
| C(5T)-C(6T) | 1.363(2) | | |
| C(7T)-C(8T) | 1.432(3) | | |
| C(7T)-C(9T) | 1.434(3) | | |
| C(10T)-C(11T) | 1.428(3) | | |
| C(10T)-C(12T) | 1.429(3) | | |
| | | | |
| C(7T)-C(1T)-C(6T) | 120.91(17) | C(16T)-C(13T)-C(14T) | 121.58(17) |
| C(7T)-C(1T)-C(2T) | 121.01(17) | C(16T)-C(13T)-C(15T) | 120.81(16) |
| C(6T)-C(1T)-C(2T) | 118.08(16) | C(14T)-C(13T)-C(15T) | 117.60(15) |
| C(3T)-C(2T)-C(1T) | 120.88(17) | C(15TA)-C(14T)-C(13T) | 121.01(17) |
| C(2T)-C(3T)-C(4T) | 121.28(18) | C(14TA)-C(15T)-C(13T) | 121.38(16) |
| C(10T)-C(4T)-C(3T) | 120.92(17) | C(13T)-C(16T)-C(18T) | 120.45(17) |
| C(10T)-C(4T)-C(5T) | 121.24(17) | C(13T)-C(16T)-C(17T) | 123.14(17) |
| C(3T)-C(4T)-C(5T) | 117.76(16) | C(18T)-C(16T)-C(17T) | 116.31(16) |
| C(6T)-C(5T)-C(4T) | 121.09(17) | N(6T)-C(17T)-C(16T) | 177.4(2) |
| C(5T)-C(6T)-C(1T) | 120.86(17) | N(5T)-C(18T)-C(16T) | 178.3(2) |
| C(1T)-C(7T)-C(8T) | 121.56(17) | | |
| C(1T)-C(7T)-C(9T) | 122.02(17) | | |
| C(8T)-C(7T)-C(9T) | 116.42(16) | | |
| C(4T)-C(10T)-C(11T) | 121.17(17) | | |
| C(4T)-C(10T)-C(12T) | 121.72(18) | | |
| C(11T)-C(10T)-C(12T) | 117.00(17) | | |

[a] All atoms with label **A** are obtained from the base ones by the symmetry operations -x+1,-y+2,-z+1.

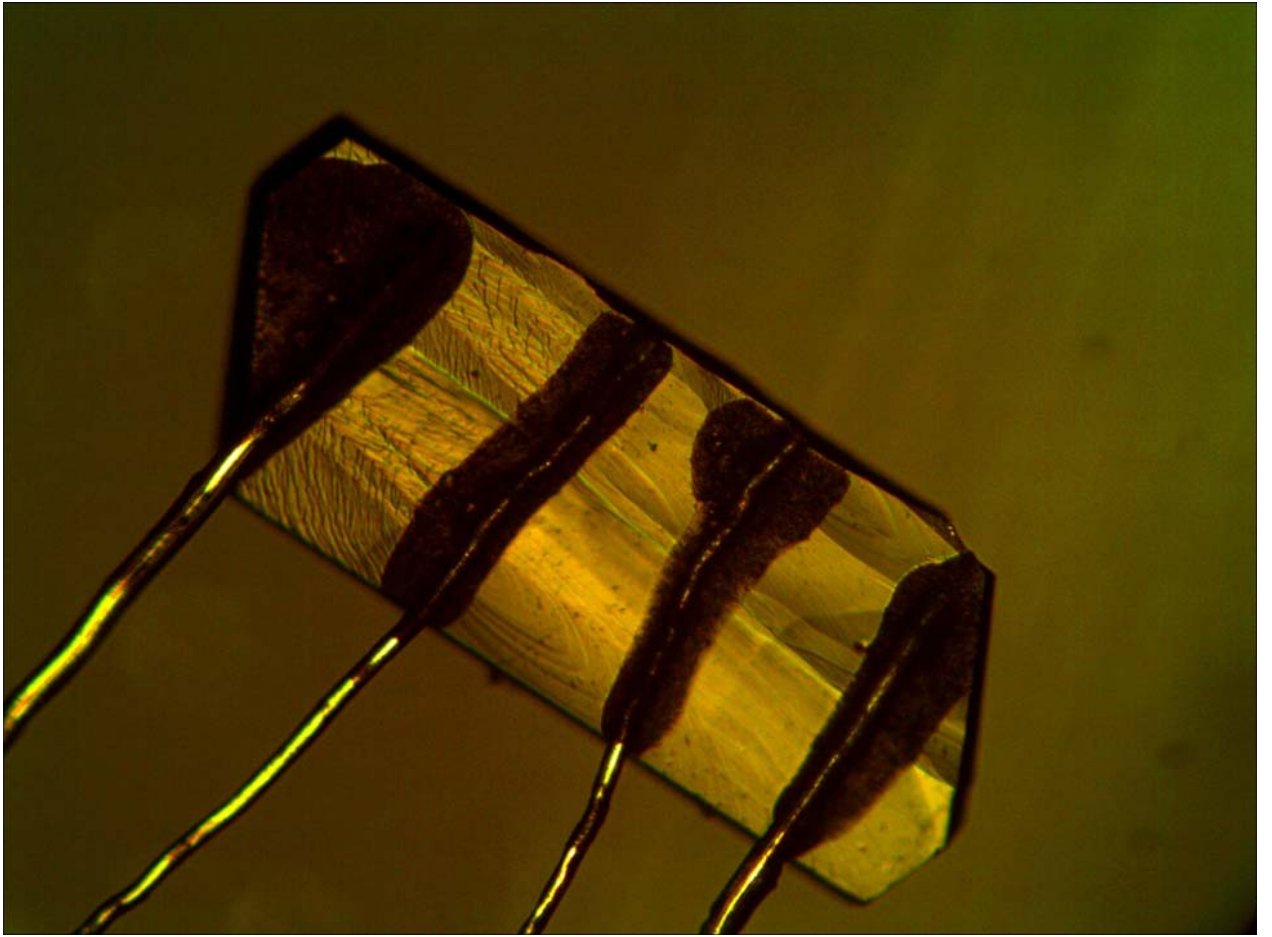

**Fig. S8.** The placement of the electrodes on the crystal **I**

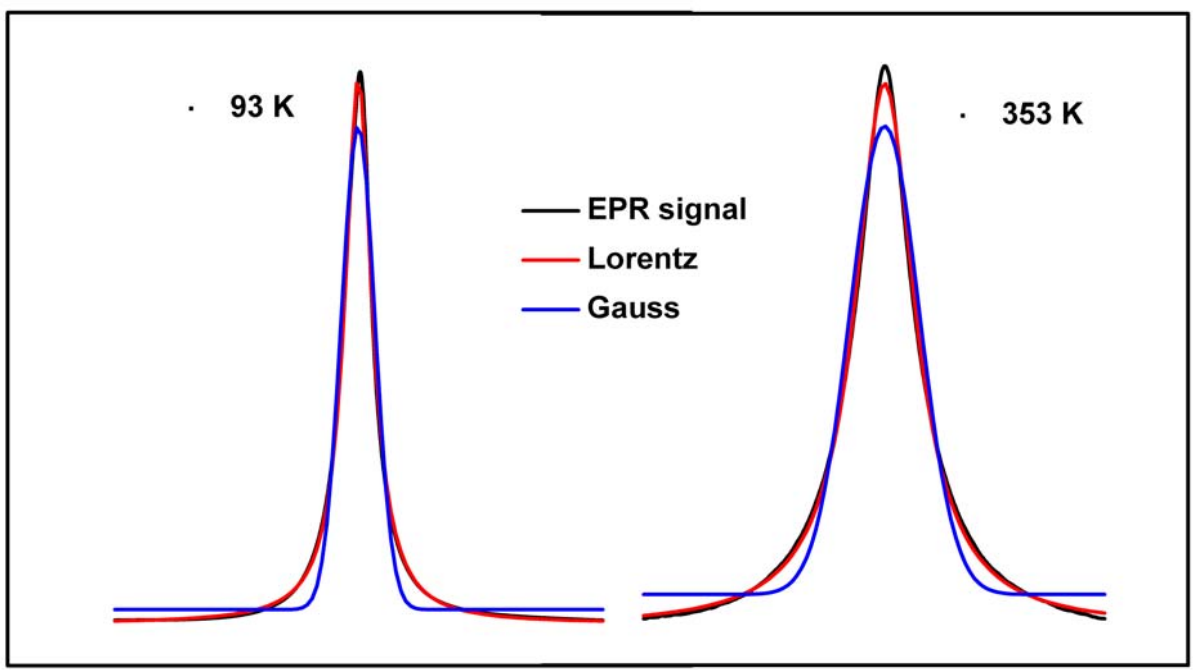

**Fig. S9.** Comparison of the central component of the EPR signal (black dots) with Lorentzian (red dots) and Gaussian (blue dots) signals at 93 K and 353 K.

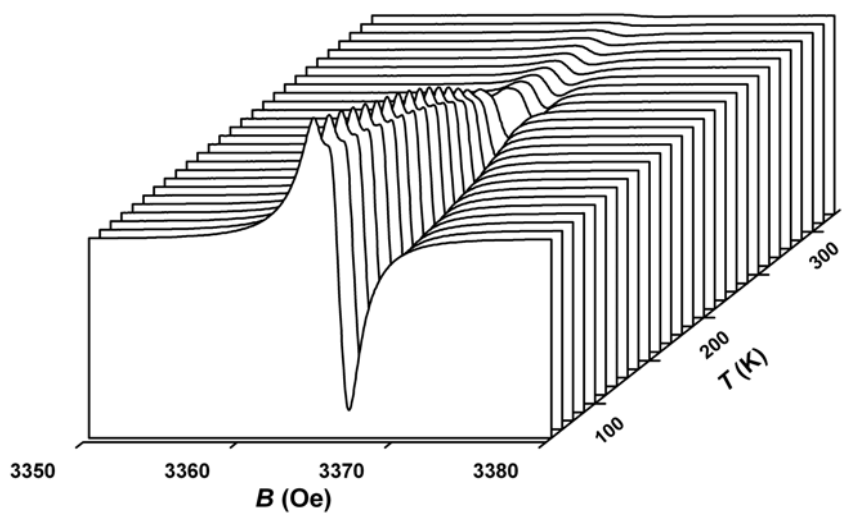

**Fig. S10.** Temperature evolution of the EPR lineshape.